\documentclass[floatfix,twocolumn,nofootinbib,superscriptaddress,a4paper,preprintnumbers]{revtex4-1}
\usepackage{graphicx}

\usepackage{amsmath}
\usepackage{epsfig,bbm}
\usepackage{url}

\usepackage{graphicx}
\usepackage{setspace}
\usepackage{amssymb}
\usepackage{listings}
\usepackage{epstopdf}
\usepackage{float}
\usepackage{ulem}
\usepackage{color}
\usepackage{amsmath}
\usepackage{dsfont}
\usepackage{slashed}

\usepackage{amsmath}

\def\cm{\rm cm}

\begin{document}
\hfill CERN-PH-TH-2015-029 \vspace{0.2cm}

\title{The Supersymmetric Standard Models with a Pseudo-Dirac Gluino from Hybrid $F-$ and
$D-$Term Supersymmetry Breakings }

\author{Ran Ding}
\affiliation{Center for High-Energy
Physics, Peking University, Beijing, 100871, P. R. China}
\author{Tianjun Li}
\affiliation{State Key Laboratory of Theoretical Physics
and Kavli Institute for Theoretical Physics China (KITPC),
Institute of Theoretical Physics, Chinese Academy of Sciences,
Beijing 100190, P. R. China}
\affiliation{School of Physical Electronics,
University of Electronic Science and Technology of China,
Chengdu 610054, P. R. China}
\author{Florian Staub}
\affiliation{Theory Division, CERN 1211 Geneva 23, Switzerland}
\author{Chi Tian}
\affiliation{School of Physical Electronics,
University of Electronic Science and Technology of China,
Chengdu 610054, P. R. China}
\author{Bin Zhu}
\affiliation{State Key Laboratory of Theoretical Physics
and Kavli Institute for Theoretical Physics China (KITPC),
Institute of Theoretical Physics, Chinese Academy of Sciences,
Beijing 100190, P. R. China}
\affiliation{Institute of Physics Chinese Academy of sciences, Beijing 100190, P. R. China}

\begin{abstract}

We propose the Supersymmetric Standard Models (SSMs) with a pseudo-Dirac gluino from hybrid $F-$ and
$D-$term supersymmetry (SUSY) breakings. Similar to the SSMs before the LHC, all the
supersymmetric particles in the Minimal SSM (MSSM) obtain the SUSY breaking
soft terms from the traditional gravity mediation and have masses within about 1 TeV except gluino.
To evade the LHC SUSY search constraints, the gluino also has a heavy Dirac mass above 3 TeV
from $D-$term SUSY breaking. Interestingly, such a heavy Dirac gluino mass will not induce the
electroweak fine-tuning problem. We realize such SUSY breakings via an anomalous $U(1)_X$ gauge
symmetry inspired from string models. To maintain the gauge coupling unification and increase
the Higgs boson mass, we introduce extra vector-like particles. We study the viable parameter
space which satisfies all the current experimental constraints, and present a concrete benchmark
point. This kind of models not only preserves the merits of pre-LHC SSMs such as naturalness,
dark matter, etc, but also solves the possible problems in the SSMs with Dirac gauginos
due to the $F$-term gravity mediation.

\end{abstract}

\maketitle

{\bf Introduction}---It is well-known that the weak scale supersymmetry (SUSY) is
the most promising extension for physics beyond
the Standard Model (SM)~\cite{Martin:1997ns}. It provides a well-motivated and complete
framework to understand the basic questions of TeV-scale physics: the gauge hierarchy problem
is solved naturally, the lightest supersymmetric particle (LSP) such as neutralino can be
a dark matter candidate, and gauge coupling unification can be realized, etc.
The gauge coupling unification strongly suggests the Grand Unified Theories (GUTs),
and only the superstring theory may describe the real world. Thus, the supersymmetric SM (SSM) is also a bridge
between the low energy phenomenology and high-energy fundamental physics.

However, the discovered SM-like Higgs boson with a mass around 125~GeV~\cite{Chatrchyan:2012ufa,Aad:2012tfa}
is a little bit too heavy in the Minimal SSM (MSSM) since it requires the multi-TeV top squarks with
small mixing or TeV-scale top squarks with large mixing~\cite{Carena:2011aa}.
Also, there exist strong constraints on the SSMs from the LHC SUSY searches.
For example, the gluino mass $m_{\tilde g}$ and first two-generation squark mass $m_{\tilde q}$
should be heavier than about 1.7 TeV if they are roughly degenerate $m_{\tilde q} \sim m_{\tilde g}$,
and the squark mass $m_{\tilde q}$  is heavier than about 850~GeV
for $  m_{\tilde g} \gg m_{\tilde q}$~\cite{Aad:2014wea}.
Therefore, the naturalness of the SSMs is challenged.

The basic idea to lift Higgs mass without threatening the hierarchy problem is the introduction
of additional tree-level contributions~\cite{Erler:2002pr, Ellwanger:2009dp,Ellwanger:2006rm,Ma:2011ea,Zhang:2008jm,Hirsch:2011hg,Ross:2011xv}. To escape the LHC SUSY search constraints, there are quite a few proposals:
natural SUSY~\cite{Dimopoulos:1995mi,Cohen:1996vb}, compressed SUSY~\cite{LeCompte:2011cn, LeCompte:2011fh, Dreiner:2012gx},
stealth SUSY~\cite{Fan:2011yu}, heavy LSP SUSY~\cite{Cheng:2014taa},
$R$-parity violation~\cite{Csaki:2011ge,Csaki:2013jza},
supersoft SUSY~\cite{Fox:2002bu,Benakli:2008pg,Benakli:2010gi,Kribs:2012gx,Benakli:2012cy,Kribs:2013oda,Bertuzzo:2014bwa,Benakli:2014cia,Diessner:2014ksa, Nelson:2015cea}, etc. Here, we would like to point
out that all the sparticles in the SSMs can be within about 1 TeV as long as
the gluino is heavier than 3 TeV, which is obviously an simple modification to
the SSMs before the LHC. Also, such a heavy gluino will not induce the electroweak
fine-tuning problem if it is (pseudo-)Dirac like the supersoft SUSY. However, there exists some problems
for supersoft SUSY with Dirac gauginos:
$\mu$ problem can not be solved via the Giudice-Masiero (GM) mechanism~\cite{Giudice:1988yz}, the D-term contribution
to the Higgs quartic coupling vanishes, the right-handed slepton may be the LSP, and the scalar components
of the adjoint chiral superfields might be tachyonic and then break the SM gauge symmetry, etc~\cite{Fox:2002bu}. The
first three problems can be solved in the $F-$term gravity mediation, while the last problem was solved
recently~\cite{Nelson:2015cea}. Therefore, we will propose the SSMs with a pseudo-Dirac gluino from hybrid
$F-$ and $D-$term SUSY breakings. To be concrete, all the sparticles in the MSSM
 obtain SUSY breaking soft terms from the traditional gravity mediation, and only gluino receives
extra Dirac mass from the $D-$term SUSY breaking. Especially, all the MSSM sparticles except gluino
can be within about 1 TeV as the pre-LHC SSMs. The merits of this proposal are: keeping the good
properties of pre-LHC SSMs (naturalness, as well as explanations for the dark matter and muon anomalous magnetic moment,
etc), evading the LHC SUSY search constraints, and solving the problems in supersoft SUSY via $F$-term
gravity mediation.  We show that such SUSY breakings can be realized by an anomalous $U(1)_X$ gauge
symmetry inspired from string models. To achieve the gauge coupling unification and increase the Higgs boson mass, we will
introduce vector-like particles. We shall discuss the low energy phenomenology, and the detailed studies will
be given elsewhere~\cite{DLSTZ}.

{\bf Model Building}---In order to generate the Dirac gluino mass, a chiral superfield $\Phi$ in the adjoint
representation of $SU(3)_C$ is needed. To maintain the gauge coupling unification and lift the Higgs boson mass,
we need to introduce some extra vector-like particles. To avoid the Landau pole for the SM gauge couplings
below the GUT scale, we only have two kinds of models: $\Delta b=3$ and $\Delta b=4$ where $\Delta b$ is 
 the uniform contribution to the one-loop beta functions of the SM gauge couplings from all the new particles. 
The additional vector-like particles and their quantum numbers in the supersymmetric SMs with $\Delta b =3$
and $\Delta b =4$ are given in Tables \ref{tab:Db=3} and \ref{tab:Db=4}, respectively. We will study the model
with $\Delta b=3$ elsewhere (For Dirac gaugino case, see  Ref.~\cite{Benakli:2014cia}.). Here, we shall consider
the model with $\Delta b=4$. 
In this model, the $SU(2)_L\times U(1)_Y$ Dirac gaugino masses are forbidden, and
the neutrino masses and mixings can be generated via Type II seesaw mechanism~\cite{Konetschny:1977bn}.
From the string model building point of view,
we usually do not have the vector-like particles $T_+$ and $T_-$ since they
arise from an symmetric $\mathbf{15}$ representation of $SU(5)$. Interestingly,
the symmetric $\mathbf{15}$ representation of $SU(5)$ or flipped $SU(5)$ can indeed be obtained 
in the Type IIA orientifold on $\mathbf{T^6/(Z_2\times Z_2)}$ with intersecting
D6-branes~\cite{Cvetic:2002pj, Chen:2006ip}. The alternative way to get $T_+$ and $T_-$ is to embed $SU(2)_L$
into a diagonal gauge group of $SU(2)_A\times SU(2)_B$, which was done in 
a particular $Z_3\times Z_3$ orbifold of the heterotic string~\cite{Langacker:2005pf}. In this case,
the Type II seesaw mechanism can be realized as well~\cite{Langacker:2005pf}.

\begin{table}[h]
\begin{tabular}{|c|c|c|c|}
\hline
Particles & Quantum Numbers  & Particles  & Quantum Numbers \\
\hline
$\Phi$ & $(\mathbf{8}, \mathbf{1}, \mathbf{0})$ & $T$ & $(\mathbf{1}, \mathbf{3}, \mathbf{0})$ \\
\hline
$XL$ & $(\mathbf{1}, \mathbf{2}, \mathbf{-1/2})$ & $XL^c$ & $(\mathbf{1}, \mathbf{2}, \mathbf{1/2})$ \\
\hline
$XE_i$ & $(\mathbf{1}, \mathbf{1}, \mathbf{-1})$ & $XE^c_i$ & $(\mathbf{1}, \mathbf{1}, \mathbf{1})$ \\
\hline
$S$ & $(\mathbf{1}, \mathbf{1}, \mathbf{0})$ & & \\
\hline
\end{tabular}
\caption{The extra vector-like particles and their quantum numbers in the supersymmetric SM with $\Delta b =3$. 
Here, $i=1,~2$, and we do not have to introduce $S$ except for Dirac gaugino case 
since it is an SM singlet.}
\label{tab:Db=3}
\end{table}

\begin{table}[h]
\begin{tabular}{|c|c|c|c|}
\hline
Particles & Quantum Numbers  & Particles  & Quantum Numbers \\
\hline
$\Phi$ & $(\mathbf{8}, \mathbf{1}, \mathbf{0})$ &  & \\
\hline
$XD$ & $(\mathbf{3}, \mathbf{1}, \mathbf{-1/3})$ & $XD^c$ & $(\mathbf{\bar 3}, \mathbf{1}, \mathbf{1/3})$ \\
\hline
$T_+$ & $(\mathbf{1}, \mathbf{3}, \mathbf{1})$ & $T_-$ & $(\mathbf{1}, \mathbf{3}, \mathbf{-1})$ \\
\hline
\end{tabular}
\caption{The extra vector-like particles and their quantum numbers in the supersymmetric SM with $\Delta b =4$. }
\label{tab:Db=4}
\end{table}

Comparing to the MSSM, the new superpotential terms with universal vector-like particle mass are
\begin{align}
W&=  M_V (T_+ T_- + XD^c XD) +\lambda H_u T_- H_u + \lambda' H_d T_+ H_d \, ,\nonumber
\label{eq:DiracTMSSM}
\end{align}
where $H_d$ and $H_u$ are the MSSM Higgs fields. The $\lambda$ and $\lambda'$ terms
will respectively give positive and negative contributions to the Higgs boson mass via the non-decoupling
effects. Although with both terms we can still get the Higgs boson with mass around 125 GeV easily,
 to simplify the discussions we shall neglect the $\lambda'$ term in the following. 
The corresponding SUSY breaking soft terms are
\begin{align}
-\mathcal{L}_{soft}&= B_T T_- T_+ +B_D XD^c XD +T_{\lambda} H_u T_- H_u\nonumber\\
                   & +M_{D} G \Phi + {\rm h.c.} +  \tilde \phi^\dagger m_{\tilde\phi}^2 \tilde\phi,
\end{align}
where $B_{\mu,T,D}$ are bilinear soft terms,  $m_{\tilde\phi}^2$ are soft scalar masses,
and  $M_{D}$ is the Dirac gluino mass.

{\bf SUSY Breaking}---To realize the hybrid $F-$ and $D-$term SUSY breakings,
we shall consider the anomalous $U(1)_X$ gauge symmetry
inspired from string models~\cite{Dvali:1996rj}. Unlike Ref.~\cite{Dvali:1996rj},
we introduce two SM singlet fields $S$ and $S'$ with $U(1)_X$ charges $0$ and $-1$, and assume
that all the SM particles and vector-like particles
are neutral under $U(1)_X$. In general, there could exist other exotic particles $Q^X_i$ with
$U(1)_X$ charges $q^X_i$ from any real string compactification. Thus, the $U(1)_X$  $D$-term potential is
\begin{eqnarray}
V_D= {g_X^2 \over 2}D^2 = {g_X^2 \over 2}\left(\sum_i q^X_i|Q^X_i|^2 - |S'|^2+\xi\right)^2\, ,
\label{D-Potential}
\end{eqnarray}
where for example, in the heterotic string compactification~\cite{Cvetic:1998gv},
the Fayet-Iliopoulos term
is given by
\begin{eqnarray}
\xi = {g_X^2{\rm Tr}{ q^X} \over 384\pi^2}M_{\rm Pl}^2\, ,
\end{eqnarray}
where $M_{\rm Pl}$ is the reduced Planck scale.

To achieve gravity mediation, we consider the following superpotential from the instanton effect
which breaks $U(1)_X$
\begin{eqnarray}
W_{\rm Instanton} &=& M_I S S'~.~
\end{eqnarray}
This is the key difference between our scenario and that in Ref.~\cite{Dvali:1996rj} where the superpotential
is $U(1)_X$ invariant and then one can not realize
the traditional gravity mediation. Minimizing the potential, we obtain
\begin{eqnarray}
&& \langle S \rangle = 0~,~~\langle S' \rangle^2
= \xi- {M_I^2 /g_X^2}~,~~ \langle F_{S'}\rangle = 0~,~\\
&& \langle F_{S}\rangle = M_I\sqrt{\xi- {M_I^2 /g_X^2}}~,~~
 \langle D\rangle ={M_I^2}/{g_X^2}\, .
\end{eqnarray}
Because $S$ is neutral under $U(1)_X$, the traditional gravity mediation can be realized
via the non-zero $F_S$. The Dirac mass for gluino/$\Phi$ and soft scalar
masses for $\Phi$ and $T_{+/-}$ can be generated respectively via the following operators~\cite{Nelson:2015cea}
\begin{eqnarray}
\int d^2\theta\left( \frac{{\overline D}^2 D^\alpha V'}{M_*} W_{3, \alpha} \Phi+
\frac{{\overline D}^2(D^{\alpha} V' D_{\alpha} X' )}{M_*} X''\right) ~,~\,
\end{eqnarray}
where we neglect the coefficients for simplicity, $X'$ and $X''$ can
both be $\Phi$ as well as respectively be $T_{+/-}$ and $T_{-/+}$,
and $M_*$ can be the reduced
Planck scale for gravity mediation or the effective messenger scale.
In addition, like the above second kind of operators,
we can have the similar operators for  $H_d$/$H_u$ and $XD^c/XD$.
Although such operators for $XD^c/XD$ are fine, the operators for 
$H_d$/$H_u$ must be suppressed. Otherwise, we will not have electroweak symmetry breaking
due to large soft masses for $H_d$ and $H_u$. For simplicity,
we shall assume that these operators are suppressed due to the localizations of the particles in the
extra space dimensions in Type IIA/B string constructions, or the suppressed couplings
with messengers. To be concrete, in the Type IIA orientifold on $\mathbf{T^6/(Z_2\times Z_2)}$ with intersecting
D6-branes, all the particles except vector multiplets arise from the intersections of the D6-branes.
The Yukawa couplings in the intersecting D-brane worlds arise from open string world-sheet
instantons that connect three D-brane intersections~\cite{Aldazabal:2000cn, Chen:2007px, Chen:2007zu}. 
For a given triplet of intersections, the
minimal world-sheet action, which contributes to the Yukawa coupling, is weighted by a factor
$e^{-A_{abc}}$, where $A_{abc}$ is the world-sheet area of the triangles bounded by the branes $a$, $b$, and
$c$~\cite{Aldazabal:2000cn, Chen:2007px, Chen:2007zu}. 
Similar results hold for the four D-brane intersections (four-point interactions)~\cite{Chen:2008rx}
as well as the ${\rm E}_2$ instanton effects~\cite{Blumenhagen:2006xt, Ibanez:2006da}. 
Therefore, such operators for  $H_d$/$H_u$ and $XD^c/XD$
can be suppressed easily by adjusting the worldsheet areas due to the exponential suppressions. 
On the other hand, even if we do not consider
the explanations from the Type IIA/B string constructions, the fine-tuning measures for these coupling
hiearchies are about 25-70 since the soft masses for $\Phi$ and $T_{+/-}$ can be about 3-5 TeV. Such fine-tuning
measures are similar to the following SUSY electroweak fine-tuning.

Let us consider two cases for $M_*$: (i) We choose $M_*=M_{\rm Pl}$,
$M_I=10^8 ~{\rm GeV} $, ${\rm Tr} q^X=2$, and $g_X=10^{-3}/a$ with $a$ a real number.
So we get $D=10^{22}/a^2~{\rm GeV}^2$ and $F_S=5.5a\times 10^{21}~{\rm GeV}^2$.
 For $a=2^{-1/2}$, we have $D/F_S=5.1$, {\it i.e.}, the Dirac gluino mass and
the soft scalar masses of $T_{+/-}$ and $\Phi$ are about 5.1 times larger
than the gravity mediation via $F_S$. This case may be realized in Type IIA/B
compactifications with the D-branes wrapping the large cycles but not
in the heterotic string compactifications since $g_X$ is small.
(ii) We choose $M_I=1.25\times 10^5~{\rm GeV}$, ${\rm Tr} q^X=2$, and $g_X=0.8$,
which may be realized in heterotic string as well. So we have
 $D=2.44\times 10^{10}~{\rm GeV}^2$ and $F_S=5.5\times 10^{21}~{\rm GeV}^2$.
Thus, we need the effective messenger scale $M_*$ around $10^6$~GeV to realize the relatively heavy
masses for Dirac gluino and scalar components of  $T_{+/-}$ and $\Phi$.
In our model, the vector-like particles like $XD^c/XD$ can be messengers.

\begin{table*}[hbt]
\begin{tabular}{|c|c|c|c|c|c|c|c|c|c|c|c|c|c|}
\hline
$\tan\beta$ & $\lambda_{\text{eff}}$ & $\mu$     & $B_\mu$       & $M_1$        & $M_2$     & $M_D$   & $m_{\tilde{Q},1\&2}$  & $m_{\tilde{Q},3}$ &  $m_{\tilde{L},1\&2}$  & $m_{\tilde{L},3}$ & $m_{\Phi}$  \\
\hline
$[5,60]$    & $[0.1,0.7]$            & $[0.3,1]$ & $[10^{-3},1]$ & $[0.01,0.1]$ & $[0.5,1]$ & $[3,5]$ & $[0.8,0.9]$         & $[0.4,0.7]$       &  $[0.1,0.5]$          & $[0.07,0.16]$ & $[\sqrt{3},\sqrt{5}]$ \\
\hline
\end{tabular}
\caption{The input parameter ranges or values used in our scans. All the mass parameters are given in
appropriate power of TeV.
Here, $M_i$ are gaugino masses, $\mu$ is the bilinear Higgs mass in the superpotential and $B_{\mu}$ is
the corresponding soft mass.
We consider the universal scalar mass for the left- and right-handed
 squarks (sleptons) $\tilde{Q} \in \{\tilde{q},\tilde{d},\tilde{u}\}$ ($\tilde{L} \in \{\tilde{l},\tilde{e}\}$)
and the degenerated first and second generations.
We choose $M_3 = 0.6$~TeV and the vanishing trilinear soft terms for three generations.}
\label{tab:ranges}
\end{table*}

\begin{figure} [tp]
\begin{center}
\includegraphics[width=0.45\textwidth]{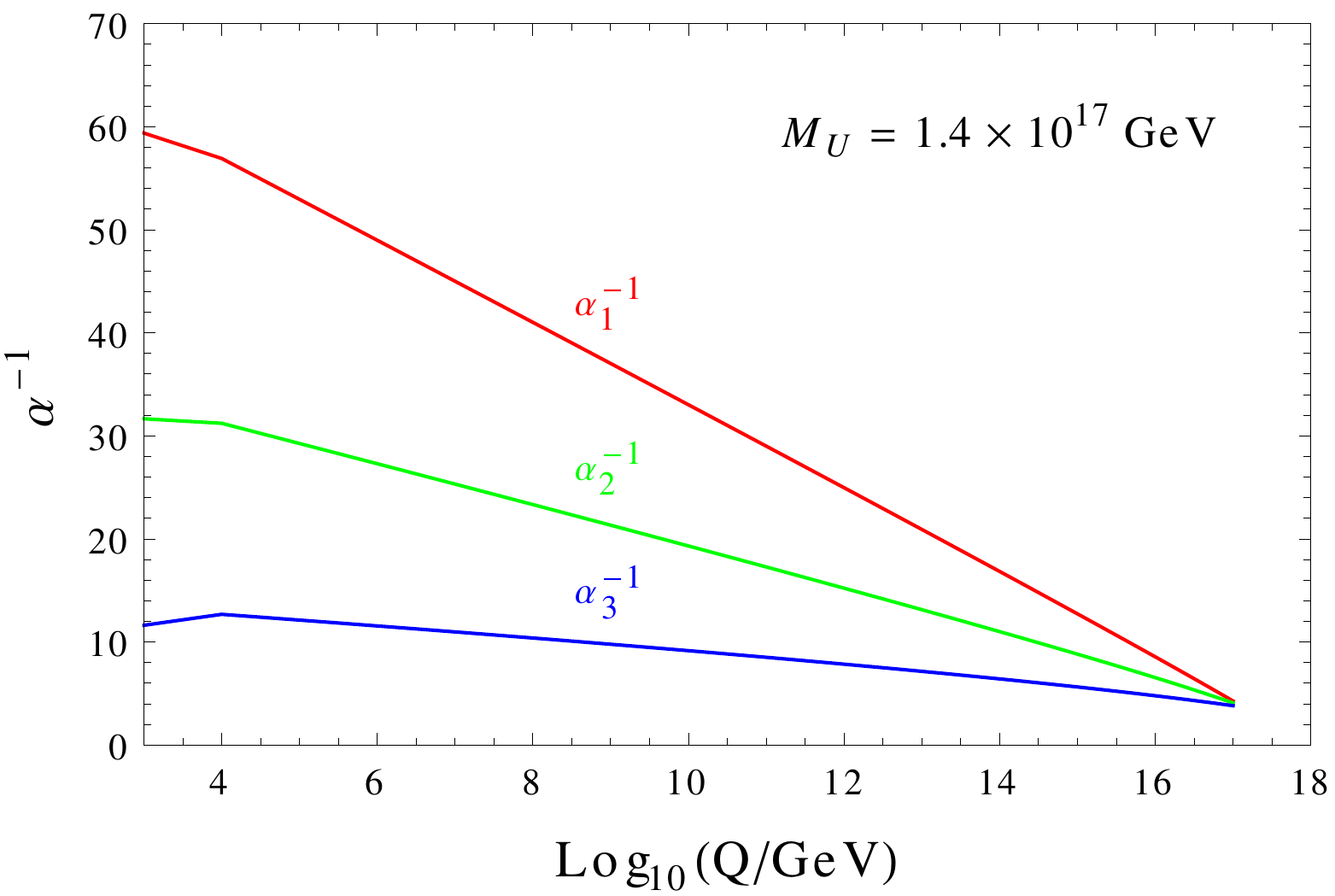}
\end{center}
\caption{Gauge coupling unification for $M_V=M_D=5$~TeV. }
\label{fig:GUT}
\end{figure}

{\bf Phenomenology Study}---First, with two-loop renormalization group equations (RGEs)
for gauge couplings and two-loop RGEs for Yukawa couplings~\cite{Martin:1993zk,Goodsell:2012fm},
we present gauge coupling unification in Fig.~\ref{fig:GUT} for $M_V=M_D=5$~TeV,
and the GUT scale is around $10^{17}$~GeV.
To avoid the Landau pole problem for gauge couplings, we need $M_V \ge 3$~TeV and $M_D \ge 3$~TeV.
Thus, the contribution to Higgs boson mass from $\lambda H_u T_- H_u$ will be suppressed. In our model, we can have
$m_{T_+}\gg M_V$, and then there exists a non-decoupling effect as in the Dirac NMSSM~\cite{Lu:2013cta,Kaminska:2014wia}.
The Higgs boson mass is increased by
\begin{eqnarray}
\Delta m^2_h = \lambda_{\text{eff}}^2  \sin^4\beta v^2\,\,,
\end{eqnarray}
where $\tan\beta \equiv \langle H_u \rangle/\langle H_d \rangle$, and
\begin{equation}
\lambda_{\text{eff}}^2 \equiv \lambda^2(m_{T_+}^2/(M_{V}^2+m_{T_+}^2)) \, .
\end{equation}
Unlike the Dirac NMSSM, this contribution does not vanish at large $\tan\beta$ limit,
which is properly accommodated with some interesting low energy constraints
such as the following $\Delta a_{\mu}$.

{\bf Numerical Results}---For the numerical studies,  we are going to study the effective theory
at the SUSY scale after integrating out the vector-like particles.
We implement our model in the Mathematica package
{\tt SARAH}~\cite{Staub:2008uz,Staub:2010jh,Staub:2011dp,Staub:2012pb,Staub:2013tta,Porod:2014xia,Goodsell:2014bna}.
{\tt SARAH} is used in a second step to  generate the various relevant outputs necessary for our analysis:
we use the Fortran modules for {\tt SPheno}~\cite{Porod:2003um,Porod:2011nf} to calculate the mass spectra
and precision observables, and the model files for {\tt CalcHEP} \cite{Belyaev:2012qa} which are used
together with {\tt micrOMEGAs}~\cite{Belanger:2013oya,Belanger:2014vza} to calculate the dark matter
relic density and direct detection rates.

We consider all the current experimental constraints from the LEP, LHC, and B physics
experiments, etc. The Higgs mass range is taken from 123~GeV to 127~GeV. Also, the SM prediction for
the anomalous magnetic moment of the muon~\cite{Davier:2010nc, Hagiwara:2011af} has a discrepancy
with the experimental results~\cite{Bennett:2006fi, Bennett:2008dy} as follows
\begin{eqnarray}
\Delta a_{\mu}\equiv a_{\mu}({\rm exp})-a_{\mu}({\rm SM})= (28.6 \pm 8.0) \times 10^{-10}.~\,
\label{bound2}
\end{eqnarray}

\begin{figure} [t]
\begin{center}
\includegraphics[width=0.45\textwidth]{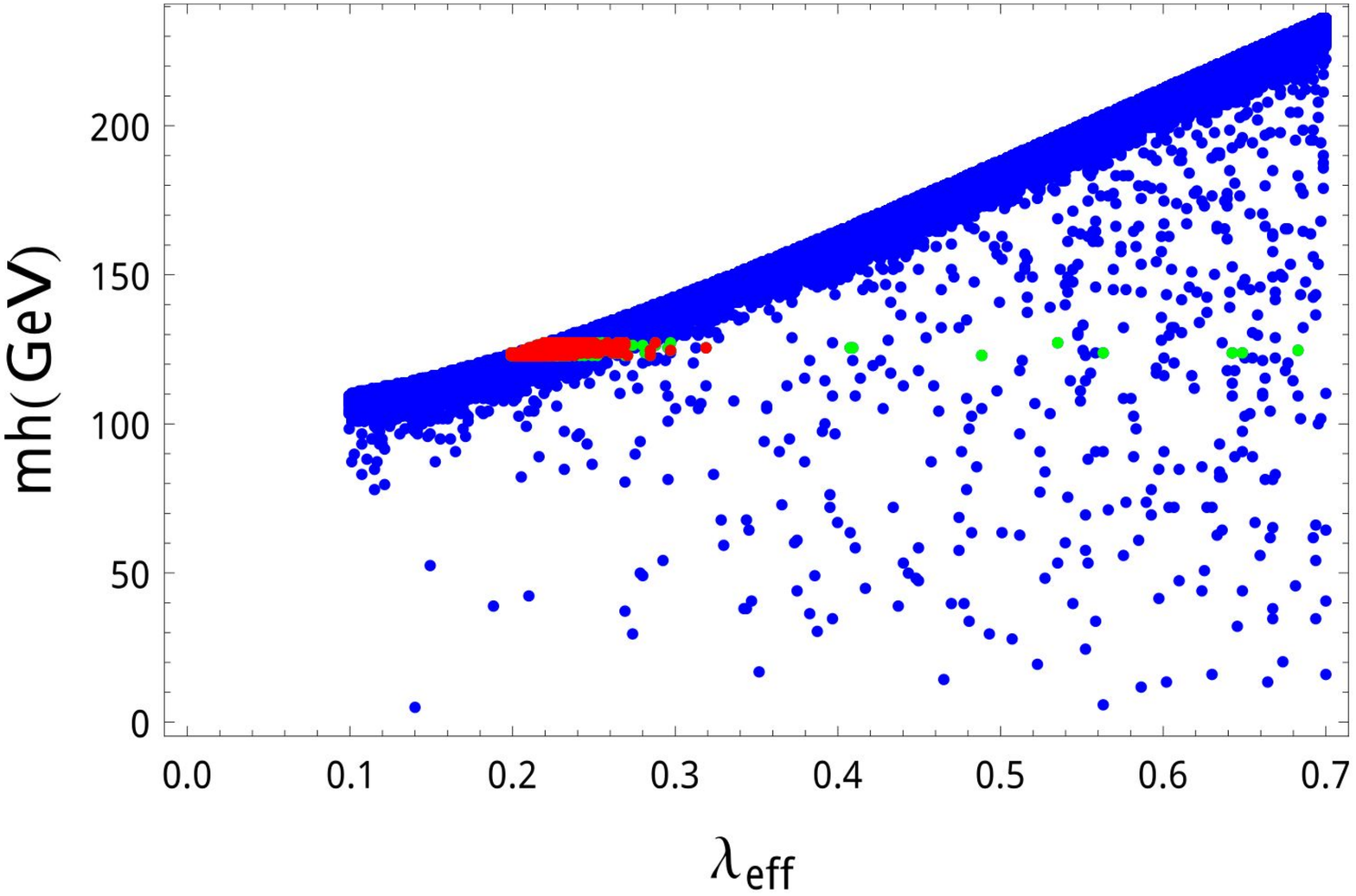}
\end{center}
\caption{The Higgs mass versus $\lambda_{\text{eff}}$.  The blue points provide
the spectra without tachyons. In addition to satisfy the Higgs mass requirement,
the green and red points have $\Delta a_{\mu}$ within 3$\sigma$ and 1$\sigma$ ranges, respectively.}
\label{fig:EWSB}
\end{figure}



\begin{table*}[ht]
\begin{center}
\begin{tabular}{|c|c|c|c|c|c|c|c|c|c|c|  } \hline
    $\widetilde{\chi}_{i}^{0}$   &$\widetilde{\chi}_{i}^{\pm}$&
$\widetilde{\nu}_e$,$\widetilde{\nu}_{\tau}$ &$\widetilde{e}_R$,
$\widetilde{e}_L$&$\widetilde{\tau}_{i}$&$\widetilde{u}_R$,$\widetilde{u}_L$
  &  $\widetilde{t}_{i}$ &$\widetilde{d}_{R}$,$\widetilde{d}_{L}$
&$\widetilde{b}_{i}$&   $h^0$      &$H^{0,\pm}/A^0$\\
\hline
    (204,446,502,561)            & (446,561)                  &
(800,257)                                    &  (802,805)
           & (211,309)            & (956,958)
  & (920,927)            &(957,962)                               &
(897,938)         &       $124.8$&$\simeq 705$   \\ \hline
\end{tabular}
\end{center}
\caption{The particle spectrum (in GeV) for a benchmark point
with pseudo-Dirac gluino masses 2927~GeV and 3470~GeV for
$\tan\beta =29$, $M_1=0.21~\text{TeV}$,
$\mu=0.5~\text{TeV}$,~$B_{\mu}=0.02~\text{TeV}^2$,
$M_2=0.5~\text{TeV}$,~$M_3=0.6~\text{TeV}$,~$M_{D}=3~\text{TeV}$,
$\lambda_{\text{eff}}=0.22$,~$m_{\Phi}=1.92~\text{TeV}$,
$m_{\tilde{Q},1\&2}=0.6~\text{TeV}$, $m_{\tilde{L},1\&2}=0.8~\text{TeV}$
$m_{\tilde{L},3}=0.26~\text{TeV}$,~$m_{\tilde{Q},3}=0.55~\text{TeV}$.
In this benchmark point, we have $\Delta_{\text {EW}}=60.4$,
$\Omega_{\widetilde{\chi}_1^0} h^2=0.1187$,
$\Delta a_{\mu} = 9.96 \times 10^{-10}$, and the
spin independent cross section $\sigma^{\rm SI}_{\widetilde{\chi}-N}= 2.85\times 10^{-46}$~$\cm^2$.}		
\label{tab:SIA1}
\end{table*}




In our scans, the input parameter ranges or values are given in Table~\ref{tab:ranges}.
In Fig.~\ref{fig:EWSB}, we present the Higgs mass versus $\lambda_{\text{eff}}$ to show
the large impact of the non-decoupling effect, in addition to the other constraints, especially
the allowed range of $\Delta a_{\mu}$. We see that for moderate values of $\lambda_{\text{eff}}$ around $0.2-0.3$,
the Higgs mass falls into the desirable range, unlike the DiracNMSSM.  Another
interesting property is that the electroweak symmetry breaking (EWSB) can be realized even in the range of
small $\mu$, which alleviates the following fine-tuning problem.

{\bf Fine-Tuning}---Because we discuss the simple low energy phenomenology here,
we consider the low energy fine-tuning measure defined
in  Ref.~\cite{Baer:2012mv} as follows
\begin{equation}
\Delta_{\text{EW}}=\frac{2}{M_Z^2}\text{max}(C_{H_d},C_{H_u},C_{\mu},C_{B_{\mu}},C_{\delta m_{H_u}^2})~,~
\label{eq:FT}
\end{equation}
where
\begin{align}
& C_{H_d}=\left|\frac{m_{H_d^2}}{\tan^2\beta-1}\right|,\, C_{H_u}=\left|\frac{m_{H_u^2}\tan^2\beta}{\tan^2\beta-1}\right|,  \\
& C_{\mu}=\left|\mu^2\right|,\,C_{B_{\mu}}=\left|B_{\mu}\right|,\\
& C_{\delta m_{H_u}^2}=\frac{(\lambda M_V)^2}{16\pi^2}\log\left(\frac{M_V^2+m^2_{T_+}}{M_V^2}\right)~.~
\end{align}
Compared to Ref.~\cite{Baer:2012mv} we have additional $C_{\delta m_{H_u}^2}$
from the triplet threshold corrections to $m_{H_u^2}$.
We find that the entire fine-tuning measure is given by $C_\mu$ while the other terms $C_{H_{d,u}}$, $C_{B_\mu}$ and
$C_{\delta m_{H_u}^2}$ are negligible. In particular, the fine-tuning measure can be as low as $50$ for the viable
parameter space, even if the threshold effects at large $M_V$ and $M_D$ are considered.
Since our MSSM sparticles except the gluino can be within about 1 TeV while gluino is Dirac, it seems
that the fine-tuning measure from high energy definition~\cite{Ellis:1986yg, Barbieri:1987fn}
will be small as well, which will be studied elsewhere.

\begin{figure} [t]
\begin{center}
\includegraphics[width=0.45\textwidth]{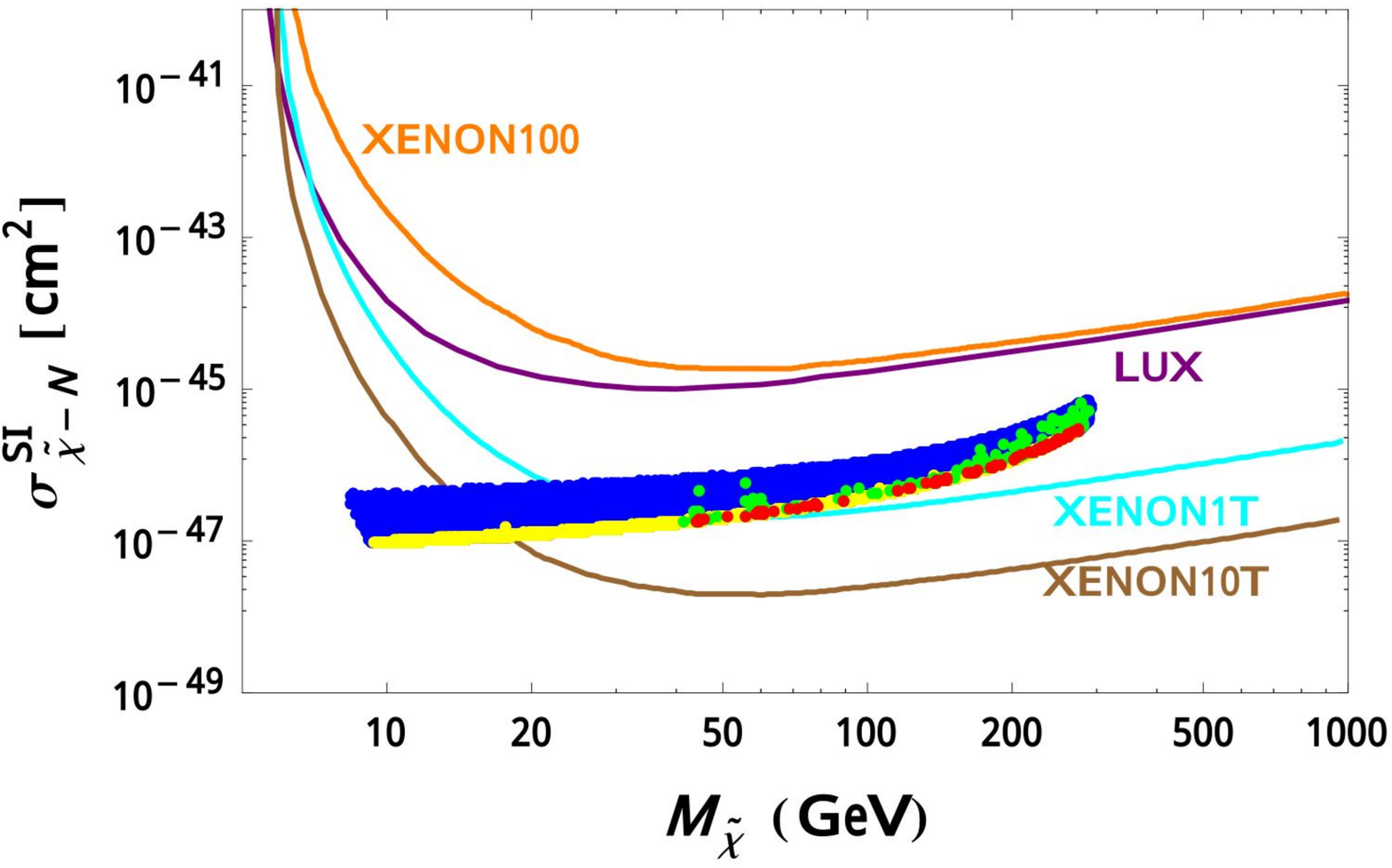}
\end{center}
\caption{The spin-independent LSP neutralino-nucleon cross section versus the LSP mass.
The blue points have the particle spectra without
tachyons. The yellow points satisfy the Higgs mass requirement
 and have $\Delta a_{\mu}$ within 3$\sigma$ range. The green points have the correct relic density
 given in Eq.~(\ref{eq:DM}). And the red points satisfy all the current constraints.}
\label{fig:directSI}
\end{figure}

{\bf Dark Matter}---For simplicity, we concentrate on the LSP neutralino-stau coannihilation scenario here.
To achieve this goal, we choose the following input parameter values or ranges:
$\mu=0.5~\text{TeV},\;B_{\mu}=0.15~\text{TeV}^2,~M_2=0.5~\text{TeV},\;
M_3=0.6~\text{TeV},\;M_{D}=3~\text{TeV},~\lambda_{\text{eff}}=0.22,\;
m_{\Phi}=2~\text{TeV},~m_{\tilde{Q},1\&2}=m_{\tilde{L},1\&2}=1~\text{TeV},\;
m_{\tilde{Q},3}=0.404~\text{TeV}$,
$5<\tan\beta<30,~ 10 ~\text{GeV}<M_1<300 ~\text{GeV},~90~\text{GeV}<m_{\tilde{L},3}<300~\text{GeV}.$
All the other parameters are taken as in Table~\ref{tab:ranges}.
 We use the relatively large values for $\mu$ and $M_2$ of 500~GeV to suppress the Higgsino and wino components
of the LSP neutralino. This reduces the direct detection rates since the coupling to the $Z$ boson
is highly reduced. Moreover,  we need a small mass splitting between the light stau and LSP neutralino to 
get an efficient coannihilation and to soften the LEP bounds on SUSY searches:
 for $m_{\tilde{\tau}_1}-m_{\tilde{\chi}_1^0}>7~\text{GeV}$ a limit of $m_{\tilde{\tau}_1}>87~\text{GeV}$ is present
while for nearly degenerated staus and neutralinos this limit becomes much weaker~\cite{Dreiner:2009ic,Beringer:1900zz}.
Finally, the fine-tuning measure is still small, $\Delta_{\text{EW}}\simeq60$.

As the preferred range for the LSP neutralino relic density,
 we consider the 2$\sigma$ interval combined range from Planck+WP+highL+BAO~\cite{Ade:2013zuv}
\begin{equation}
0.1153<\Omega_{CDM}h^2<0.1221~.~
\label{eq:DM}
\end{equation}

%

In Fig.~\ref{fig:directSI}, we show the results for spin-independent LSP neutralino-nucleon cross section.
Because the masses of the first two generations of squarks have been fixed at 1~TeV and
 the Higgsino component in the LSP is heavily suppressed, the constraints from direct detection searches are
easily evaded for all the considered points. The spin-independent cross sections are about one or
two orders of magnitude below the current best limit provided by the LUX experiment~\cite{Akerib:2013tjd}.
Especially, the points with the LSP masses above 20 (15)~GeV are within the reach of the projected XENON1T
 (XENON10T) sensitivity~\cite{xenon1t}. Also, we find that
the current constraints on spin-dependent cross sections are much weaker at the moment.
To be concrete, in Table~\ref{tab:SIA1}, we present a viable benchmark point whose MSSM particles except
gluino are within 1 TeV.

{\bf Conclusion}---We have proposed the SSMs with a pseudo-Dirac gluino from hybrid $F-$ and
$D-$term SUSY breakings, which can be achieved via an anomalous $U(1)_X$ gauge symmetry inspired from string models.
All the MSSM particles obtain the SUSY breaking soft terms from the traditional gravity mediation
and can have masses within about 1 TeV except gluino. To escape the LHC SUSY search constraints and
avoid the electroweak fine-tuning problem, the gluino also has a heavy Dirac mass above 3 TeV
from $D-$term SUSY breaking. To realize the gauge coupling unification and lift
the Higgs boson mass, we introduced extra vector-like particles. We have studied the viable parameter
space which satisfies all the current experimental constraints, and given a concrete benchmark
point.  This kind of models  keeps the merits of pre-LHC SSMs and solves the possible problems
in the supersoft SUSY.

{\bf Acknowledgements}---This research was supported in part by the
Natural Science Foundation of China under grant numbers 10821504, 11075194, 11135003, 11275246, and 11475238,
and by the National Basic Research Program of China (973 Program) under grant number 2010CB833000 (TL).



\bibliography{DiracTMSSM}
\bibliographystyle{ArXiv}

\end{document}